\documentclass[twocolumn]{aastex63}
\usepackage{amsmath}

\usepackage{xcolor, fontawesome, microtype}
\usepackage{ mathrsfs }
\usepackage[T1]{fontenc}
\usepackage{xcolor}

\newcommand{\teff}{$T_\textrm{eff}\;$}

\received{\today}
\revised{\today}
\accepted{\today}
\submitjournal{ApJL}

\shorttitle{The End of Hot Jupiter Loneliness}

\shortauthors{Zink et al. (2023)}
\begin{document}

\title{Hot Jupiters Have Giant Companions: Evidence for Coplanar High-Eccentricity Migration}

\correspondingauthor{Jon Zink}
\email{jzink@caltech.edu}

\author[0000-0003-1848-2063]{Jon K.\ Zink}
\altaffiliation{NHFP Sagan Fellow}
\affiliation{Department of Astronomy, California Institute of Technology, Pasadena, CA 91125, USA}

\author[0000-0001-8638-0320]{Andrew W.\ Howard}
\affiliation{Department of Astronomy, California Institute of Technology, Pasadena, CA 91125, USA}

\begin{abstract}
This study considers the characteristics of planetary systems with giant planets based on a population-level analysis of the California Legacy Survey planet catalog. We identified three characteristics common to hot Jupiters. First, while not all hot Jupiters have a detected outer giant planet companion ($M \sin i$ = 0.3--30 $M_{\textrm{Jup}}$), such companions are ubiquitous when survey completeness corrections are applied for orbital periods out to 40,000 days. Giant harboring systems without a hot Jupiter also host at least one outer giant planet companion per system. Second, the mass distributions of hot Jupiters and other giant planets are indistinguishable. However, within a planetary system that includes a hot Jupiter, the outer giant planet companions are at least $3\times$ more massive than the inner hot Jupiters. Third, the eccentricity distribution of the outer companions in hot Jupiter systems (with an average model eccentricity of $\langle e\rangle=0.34\pm0.05$) is different from the corresponding outer planets in planetary systems without hot Jupiters ($\langle e\rangle=0.19\pm0.02$). We conclude that the existence of two gas giants, where the outermost planet has an eccentricity $\ge0.2$ and is $3\times$ more massive, are key factors in the production of a hot Jupiter. Our simple model based on these factors predicts that $\sim$10\% of warm and cold Jupiter systems will by chance meet these assembly criteria, which is consistent with our measurement of $16\pm6\%$ relative occurrence of hot Jupiter systems to all giant-harboring systems. We find that these three features favor coplanar high-eccentricity migration as the dominant mechanism for hot Jupiter formation.

\end{abstract}

\section{Introduction}

Nearly three decades since the discovery of 51 Pegasi b, a hot Jupiter orbiting a main-sequence star \citep{may95}, the origin of these short-period ($P<10$ days) gas giants remains unresolved. In high-eccentricity migration theories, hot Jupiters are formed in more distant orbits beyond the ice line, attain high eccentricity through a dynamical interaction with other planets or stars, and then circularize and shrink their orbits through tidal interactions with the host star. Several types of dynamical interactions have been considered, including planet-planet scattering \citep{ras96,cha08}, Lidov–Kozai cycling with a distant companion \citep{wu03,naoz12,petr16,vick19}, and secular dynamics \citep{wu11,petr15}. Alternatively, hot Jupiters could form in situ \citep{bat16,boy16} or through disk-driven migration \citep{lin96}.

The detection of planetary or stellar companions in these systems may provide clarity about their formation pathways. If most hot Jupiter systems have an abundance of small nearby planetary companions, such as Kepler-730 \citep{can19}, a quiescent process is favored. Alternatively, a consistent absence of nearby planets and a massive ($M>0.3M_\textrm{Jup}$) distant ($a>1$ AU) companion might suggest a dynamically driven origin. Using direct imaging techniques, \citet{ngo16} found that less than 20\% of planetary systems with a hot Jupiter have a stellar companion close enough to incite the necessary Kozai–Lidov oscillations required to drive orbital evolution. Furthermore, the vast majority of hot Jupiters from the \emph{Kepler} survey do not show evidence for a second transiting planet in the system with an orbital period of less than 400 days (such planets could be present, but mutually inclined so that transits are not expected). Searches for periodic ephemeris variations caused by nearby planet-planet interactions have produced mixed results. Early work found a lack of such signals \citep{ste12}, suggesting that hot Jupiters orbit their host stars without additional planetary companions. Recent work using the full Kepler baseline identified two (out of the 50 hot Jupiters) that show transit-timing variations (TTVs) \citep{wu23}, suggesting that $12\%$ harbor small, nearby planets ($M<10M_{\Earth}$, $P<40$ days). Analyses of long-period RV accelerations in hot Jupiter systems found that 50--70\% have a massive planetary companion within 1--20 AU \citep{knu14,bry16}. A limitation of this work is that it relied on heterogeneous RV datasets whose baselines didn't cover the orbital periods of the proposed outer companions; extrapolation and modeling of RV ``trends'' were needed to infer the presence of outer companions. In contrast, \citet{sch16} carried out a meta-analysis of giant hosting systems using values derived in \citet{cum08} and \citet{wri12}, alongside a sample of 136 known RV systems which host a longer-period giant ($P>10$ days). This work found no difference in the occurrence of distant giants in planetary systems with or without hot Jupiters.

In this work, we analyze the planet catalog and completeness measurements from the California Legacy Survey (CLS; \citealt{ros21}). CLS has several features that make it well-suited for population studies of giant planets. The set of nearby stars in CLS was blindly selected and observed over decades (30 years in some cases) with minimum observing standards. Each star's RV time series was searched with the same algorithm, and the catalog of planets was constructed using estimates of the search completeness for each star. Analysis of the CLS planet catalog has produced measurements of the occurrence of giant planets from 0.1 to 30 AU \citep{ful21}. Our analysis of the CLS catalog in this paper focuses on planetary systems with one or more giant planets, including those with a hot Jupiter. We attempt to determine a set of factors that are unique to planetary systems with hot Jupiters to find clues about how those unusual planets formed.

\begin{figure*}
\centering \includegraphics[width=\textwidth{}]{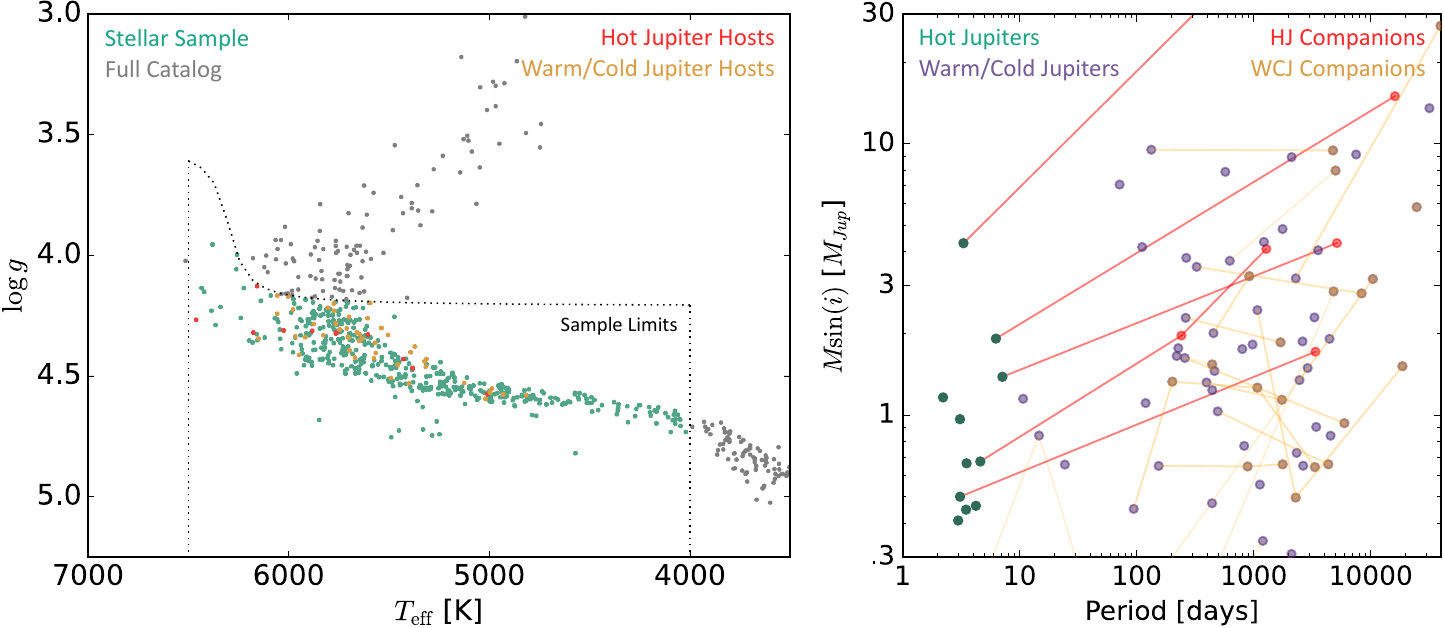}
\caption{\textbf{Left}: The stellar sample from the CLS catalog \citep{ros21} in the \teff and $\log g$ plane. The gray points show the entire sample, while the colored points represent the selected FGK dwarf population. \textbf{Right}: The sample of giant planets in this study, including hot Jupiters (teal), warm/cold Jupiters (purple), outer giant companions of hot Jupiters (red), and outer giant companions of warm/cold Jupiters (orange). The red lines connect hot Jupiters with their outer giant companions, and the orange lines connect the warm/cold Jupiter companions.
\label{fig:stellar}}
\end{figure*}

\section{Stellar and Planet Sample}
\label{sec:sample}
The current study is based on the CLS catalog (719 total stars; \citealt{ros21}). We chose to focus on main sequence FGK stars because the formation mechanisms for short-period giants around lower-mass stars may differ. For consistency with previous transit analyses (i.e., \citealt{zin23}), we selected stars with \teff ranging from 4000 K to 6500 K and 
\begin{equation}
\log g >\frac{\arctan\bigg(\frac{6300\,\mathrm{K}-T_\textrm{eff}}{67.172\,\mathrm{K}}\bigg)}{4.671}+3.876 \textrm{ dex,}
\end{equation}
based on the prescription provided in \citet{hub16}. In doing so, we retained 483 stars from the parent stellar sample (see Fig. \ref{fig:stellar}). We found a median [Fe/H] of 0.04 dex and a dispersion of 0.3 dex, consistent with the \emph{Kepler} field metallicity distribution \citep{don14}. The stellar catalog was then partitioned into two subsamples: hot Jupiter hosts (11 stars) and warm/cold Jupiter hosts (46 stars). These groups were used for occurrence measurements of their corresponding planet companion populations. 

Starting with the CLS planet catalog for our selected host star population (127 planets), we restricted our sample to only include Jovian-mass planets (0.3--30\,$M_\textrm{Jup}$) and with orbital periods of less than 40,000 days.
In total, 82 planets met the criteria. We note that these cuts did not impact the planet multiplicity of the HJ sample, as the sample retains all planetary companions in the CLS catalog. Of note is HD 120136, which harbors the highest mass ($4.3\,M_\textrm{Jup}$) hot Jupiter and has an identified sub-stellar companion with a mass of roughly 350\,$M_\textrm{Jup}$ (0.34 $M_\mathrm{Sun}$), warranting the exclusion of this system from our analysis. We also note that HD 187123 is a system with a HJ, an outer giant planet companion, and a long-term RV trend. This is the only HJ harboring system, in the CLS catalog, with an identified long-term RV trend. The unresolved RV trend is consistent with an orbit of $\sim$43,000 days and a planetary-mass object (RV amplitude $\sim5 m/s$). Since the CLS pipeline did not assign planet candidacy to this RV trend, it is not included in our sample (but the two inner giant planets are).

% We will discuss this complication further in Section ??.

We define four classes of giant planets that have masses ($M\sin i$) in the range 0.3--30\,$M_\mathrm{Jup}$: hot Jupiters (HJs; $P < 10$\, days), hot Jupiter companions (giant planets orbiting exterior to a HJ), warm/cold Jupiters (WCJs; $P = 10$--40,000\, days), and warm/cold Jupiter companions (giant planets orbiting exterior to a WCJ). There are 11 HJs and 5 HJ companions in the sample. The remaining 66 planets are WCJs (46 planets) if they represent the shortest period giant in their system, or WCJ companions (20 planets) for the corresponding outer planetary companions. Fig. \ref{fig:stellar} shows the four classes. Interestingly, all HJs with companions are less massive than their companions (see Section \ref{sec:mass} for further discussion).

\section{Modeling the Population of Giant Planets}

We modeled the occurrence distribution of giant planets as the product of three functions describing their orbital period ($P$), mass ($M$), and eccentricity ($e$): 
\begin{equation} \label{eq:occ}
\frac{ d^3n}{d \log P \: d \log M \: d e} = f\: g(P)\: q(M)\: \beta(e) .
\end{equation}
We used a broken power law $g(P; \alpha_1, P_\mathrm{br}, \alpha_2)$ to model the orbital period dependence and a single power law model $q(M; \alpha)$ to model the planet mass function. $P_\mathrm{br}$ represents the break in the power law, and the $\alpha$'s are the power law parameters. 
We used a two-parameter beta distribution $\beta(e; \gamma, \lambda)$ to model eccentricity, as suggested by \citet{hog10}.
The occurrence distribution is normalized by an overall factor $f$, which represents the number of planets per star within the range of our sample. It is important to note that the population model in Equation \ref{eq:occ} treats each parameter independently. Therefore, we cannot directly resolve parameter correlations. However, dividing the planet populations into four classes---HJs, HJ companions, WCJs, WCJ companions---allows us to indirectly investigate associations via comparisons of the models of each planet category.

\begin{figure*}
\centering \includegraphics[width=\textwidth{}]{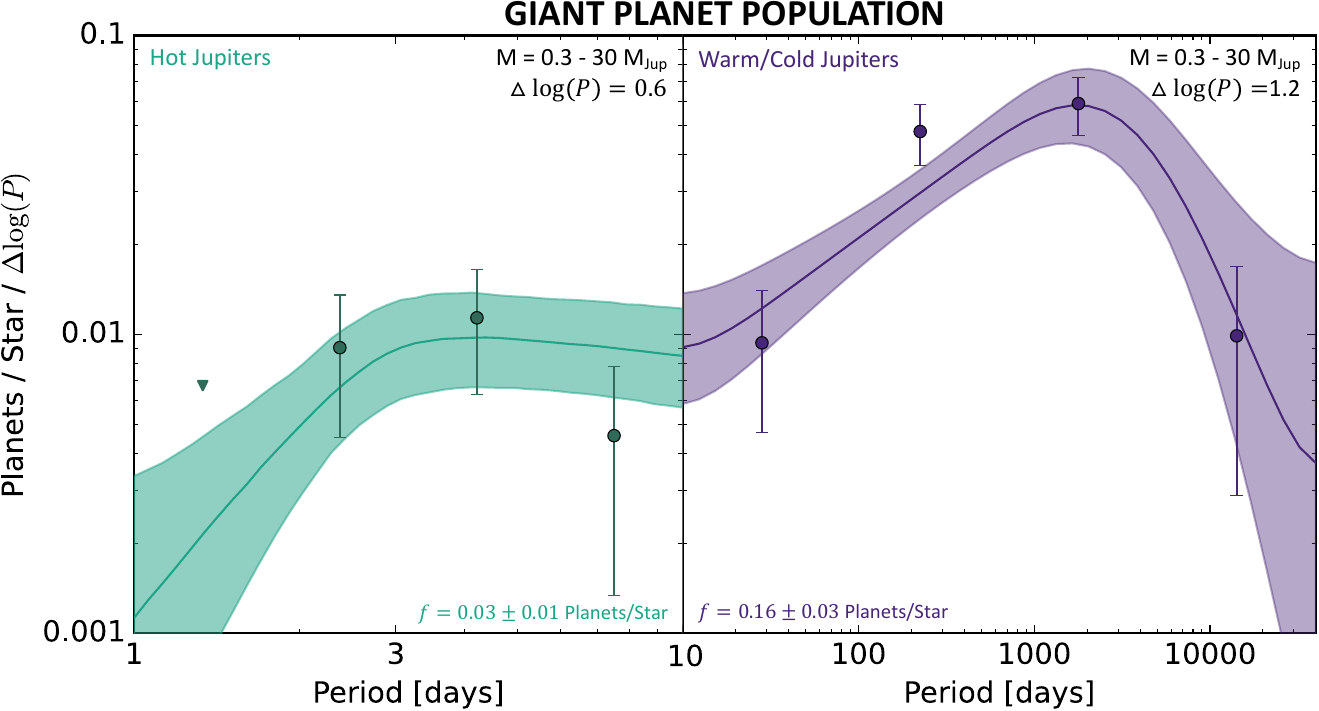}
\caption{The completeness corrected hot Jupiter (\textbf{Left}) and warm/cold Jupiter (\textbf{Right}) planet populations as a function of orbital period. The optimized broken power-law models have been provided, where the width of the colored regions represents the 68\% confidence interval. For the hot Jupiters $\alpha_1=2.1\pm1.4$, $P_\textrm{br}=3.0\pm0.8$ days, and $\alpha_2=-0.1\pm0.3$. For warm/cold Jupiters $\alpha_1=0.4\pm0.1$, $P_\textrm{br}=2800\pm880$ days, and $\alpha_2=-1.5\pm1.3$. The binned occurrence values (and $3\sigma$ upper limits, $\triangledown$) show the underlying population trends, but were not used to fit our models. Instead, we relied on CDF matching within our forward modeling software.
\label{fig:period}}
\end{figure*}

We employed a forward modeling approach to determine the underlying demographic model parameters by analyzing a simulated planet population. While we provide a brief summary of the optimization procedure here, \citet{zin20b} has a more comprehensive explanation. Our demographic analysis begins with the stellar sample and assigns fictitious planets based on the population model described in Equation \ref{eq:occ}. These simulated planet-host systems undergo a reduction in radial velocity amplitude by accounting for the $\sin i$ factor and are subject to the completeness selection effects outlined in \citet{ros21}. Our model takes into account the individual completeness map of each star, capturing any selection effects or observing strategy choices. It is also important to note that the \citet{ros21} injection/recovery survey, which provides the basis for our completeness mapping, included the full range of eccentricities and the relevant corresponding orbital parameters, capturing the reduced detection efficiency due to the high eccentricity periapsis window function. Through numerous iterations, we simultaneously optimized the model parameters, resulting in a population of simulated observables ($P$, $M$, and $e$) that match the planet sample discussed in Section \ref{sec:sample}. This process compared the $P$, $M$, and $e$ cumulative distribution functions of the simulated sample with the observed population and minimized their corresponding two-sample Anderson-Darling test statistic \citep{and52}.

\section{Period Distribution}

The period distribution encodes information about the formation and evolution of giant planets and has been the subject of numerous analyses. For example, \citet{how12} and \citet{wit10} found a break (or pile-up) in the period distribution near three days, which may be related to tidal effects or an inner formation boundary. A flat distribution or a slight dip in occurrence has been noted out to 10 days \citep{san16}, at which point the occurrence of Jupiter-size planets appears to increase \citep{wit10}. At much longer periods of $\sim$2,500 days ($\sim3.6$ AU), \citet{ful21} found another turnover in occurrence, suggestive of a drop in planet formation beyond the ice line. Fig. \ref{fig:period} shows this drop in giant planet occurrence for the widest orbits using the current sample. Overall, the shape of the giant planet occurrence distribution in Fig. \ref{fig:period} is consistent with previous studies. The total occurrence rates for our sample limits are $3\pm1\%$ for HJs and $16\pm3\%$ for WCJs orbiting FGK dwarf stars. 

%For purposes of this study, we will assume warm/cold Jupiters formed in-situ or experience dynamically cool migration. \citet{wu23} found a significant number of TTVs for these planets, suggesting more than 70\% exist in planetary systems with small planets, which would be unlikely under the pressure of significant dynamic migration.

\subsection{Hot Jupiter Companions}
Hot Jupiter companions provide insight into the dynamics (or lack thereof) that have driven these short-period giants to their current orbital locations. In particular, theories of HJ formation/migration make different predictions about the observed properties of these systems, including the necessity of outer companions and the masses and eccentricities of the planets involved. The high-eccentricity migration model predicts that outer giant companions should be common for stars hosting HJs. In this theory, the outer planet launches the inner planet into a high-eccentricity orbit. Subsequent tidal interactions between the host star and the inner planet dissipate the orbital eccentricity, leaving behind a short-period gas giant.

\begin{figure}
\centering \includegraphics[width=\columnwidth{}]{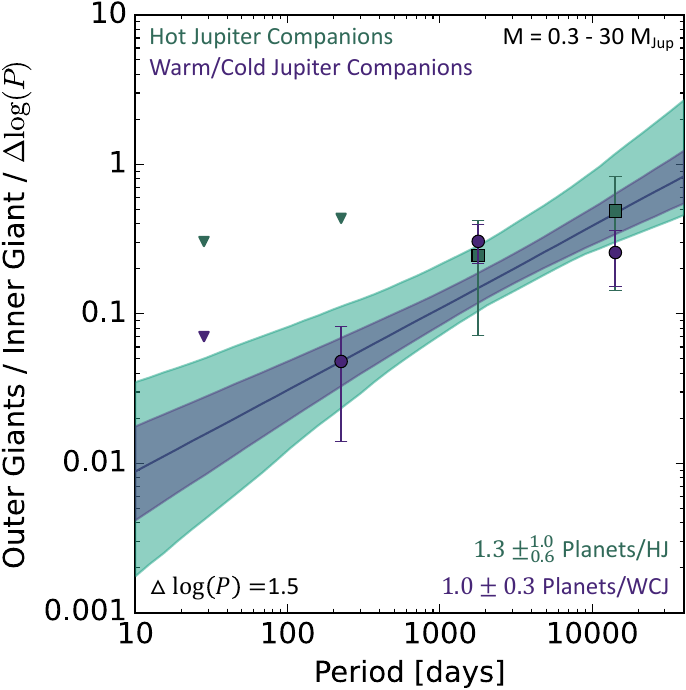}
\caption{The occurrence of an outer giant given the existence of an inner giant. The plotted points are binned occurrence values (and $3\sigma$ upper limits, $\triangledown$) showing the non-parametric population trends. The teal and purple depict the $1\sigma$ power-law models region for hot Jupiter companions ($\alpha=0.56\pm0.3$) and warm/cold Jupiter companions ($\alpha=0.54\pm0.1$) respectively.
\label{fig:compPeriod}}
\end{figure}

Fig. \ref{fig:compPeriod} shows the occurrence of companions to HJs and the WCJs. A single power law model was used, as it best replicates the period distributions. The overlap of the two bands shows that these populations have similar shapes in their occurrence distributions and similar overall occurrence levels ($1.3\pm^{1.0}_{0.6}$ planets per HJ system and $1.0\pm0.3$ planets per WCJ system, integrating the model over the 10--40,000-day interval). The steep occurrence gradient with orbital period (HJ Companions: $\alpha=0.56\pm0.3$; WCJ Companions: $\alpha=0.54\pm0.1$) explains why these outer giants have evaded detection in shorter baseline transit analyses like \citet{hua16}, who found no companions to HJs out to orbital periods of 50 days using the \emph{Kepler} sample. Our population model predicts that $2\pm2\%$ of HJs should have an outer companion within a 50-day orbital period. For the limited sample of \emph{Kepler} HJs (50 planets), only $1\pm1$ companions would be expected. However, this yield estimate is further reduced by selection effects, making it unlikely that any companions would have been detected. Even if we consider all detectable Kepler periods ($P<400$ days), we would have only expected $4\pm4$ HJ companions to exist within the period limits. Subjecting this population to transit probability selection effects again reduces the expected yield to less than one. 

Since outer giant companions are not unique to HJ systems, multiplicity must not be the only factor in the formation of HJs. In the following sections, we consider other architectural factors that contribute to HJ system formation.

\section{Mass Distribution}
\label{sec:mass}
\begin{figure*}
\centering \includegraphics[width=\textwidth{}]{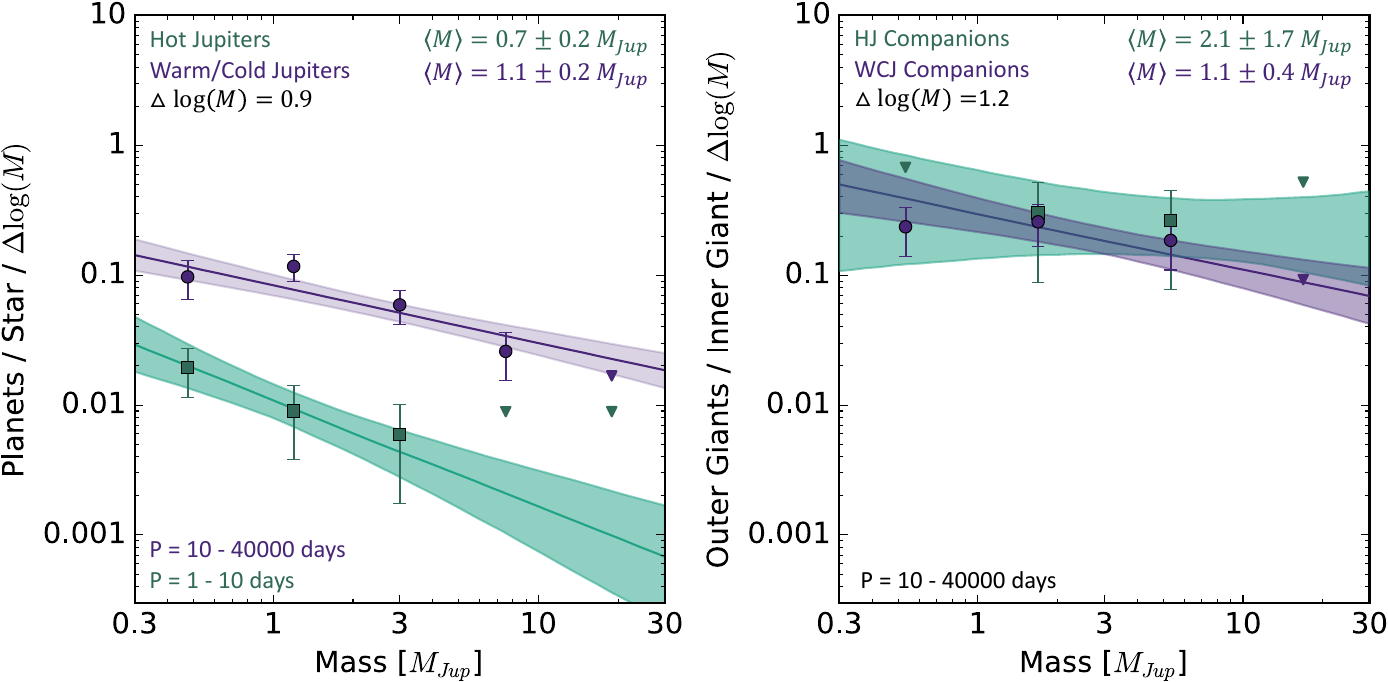}
\caption{The \textbf{Left} panel shows the underlying mass distribution of hot Jupiters ($\alpha=-0.8\pm0.3$) and warm/cold Jupiters ($\alpha=-0.4\pm0.1$). The \textbf{Right} panel depicts the corresponding mass distribution for the hot Jupiter companions ($\alpha=-0.1\pm0.4$) and the warm/cold Jupiter companions ($\alpha=-0.4\pm0.2$). For reference, we provided the expected mass values and their corresponding uncertainties as dictated by the model posteriors.
\label{fig:mass}}
\end{figure*}

Previous analyses of the HJ mass distribution noted that these planets have smaller masses compared to their warm/cold counterparts \citep{Udr03}, a potential clue about their origins. Fig. \ref{fig:stellar} is superficially consistent with this, because the highest-mass planets ($M \gtrsim 4\,M_\mathrm{Jup}$) are all WCJs ($P > 10$\,days), not HJs. However, this is an illusion because there are many more WCJs in the sample. In addition to the small number statistics at play, the low-mass WCJs are harder to detect, skewing the sample toward more massive planets. To assess the impact of this effect we forward-modeled a gas giant planet population drawn from a log-uniform mass and period distribution and found the detected WCJ population was $1.4\times$ more massive than the HJ population, despite being drawn from the same underlying mass distribution. This expected skew is roughly consistent with the 1.8 median mass ratio observed within our sample. In Fig. \ref{fig:mass} we provide the extracted underlying mass distributions for the observed HJs and WCJs. The two planet classes have approximately the same mass distribution, while HJs have a lower overall occurrence rate (the vertical offset in the left panel of Fig. \ref{fig:mass}). Specifically, our model gives power-law indices of $\alpha=-0.8\pm0.3$ and $-0.4\pm0.1$ for HJs and WCJs, respectively, indistinguishable to within $1.3\sigma$. The companion populations also have similar power-law indices, $\alpha=-0.1\pm0.4$ and $\alpha=-0.4\pm0.2$, respectively. These trends are consistent with previous giant planet mass occurrence functions (i.e., \citealt{cum08}: $\alpha=-0.3\pm0.2$ for periods within 2000 days). The observation that HJs have a mass distribution comparable to that of the outer giant planets supports the idea that they formed in a similar manner.

Although the mass distributions of HJs and WCJs are indistinguishable given the current data (see Fig. \ref{fig:stellar}), it is worth exploring whether the masses of giant planets orbiting around the same star have some relationship with each other, for example, intrasystem uniformity \citep{wei18} or increasing mass with orbital distance. To make this comparison, we first examined the five HJ systems with outer companions and found that all have substantially more massive outer companions. Excluding HD 120136, which has an outer brown dwarf companion, the mass ratios ($M_\mathrm{outer}/M_\mathrm{inner}$) are 7.8, 3.4, 3.1, and 2.9, giving a mean mass ratio of 4.2 and a typical value of $\sim$3. This striking trend can also be seen as the series of diagonal red lines that all point in the same direction in Fig. \ref{fig:stellar}. In contrast, WCJs and their nearest giant planet companions appear to have similar masses. Of the sixteen systems comprising an inner WCJ and at least one additional giant planet, only seven systems (44\%) have the more massive companion on the outside. The median mass ratio in this sample is 0.99. Furthermore, only three systems (19\%) have a mass ratio as large as the smallest value observed for HJ systems (2.9). The mass ratios of giant planets in HJ systems appear to be different from those in WCJ systems. We considered that this difference in mass ratio is due to chance. The odds of 4/4 systems being drawn randomly from the WCJ distribution with a mass ratio of at least 2.9 are $10^{-3}$ ($3.3\sigma$ significance). We also considered the possibility that this apparent intrasystem pattern is merely the outcome of selection effects. As an experiment, we forward-modeled 100,000 multi-giant configurations, drawn independently from log-uniform period and mass distributions. In cases where both planets met the WCJ criterion, the selection effects alone produced on average a $1.10\times$ more massive outer planet correlation. Similarly, in cases where the system contained a HJ and a WCJ, the selection effects on average favored a $1.09\times$ more massive outer planet. Since both configurations yield nearly identical mass ratio biases arising from completeness, the increased mass ratio observed for HJ systems appears real and not an artifact of selection effects. Therefore, HJ systems appear ordered in mass, where the outer planet is at least $3\times$ more massive. 

In our earlier comparison of the HJ and WCJ mass distributions, we found that the power-law indices $\alpha$ were indistinguishable to within $1.3\sigma$. We asserted that these system configurations appear to be drawn from the same parent population. In addition, we found that HJ systems harbor $3\times$ more massive outer companions. A critique of these two potentially competing claims is that the requirement for a $3\times$ more massive outer companion would make it less likely for high-mass HJs to arise with the same frequency dictated by random draws of the parent population, given the low probability of drawing an even higher mass companion. Therefore, the underlying HJ system mass model indices ($\alpha$) might not perfectly replicate the WCJ system models, even if the planets were drawn from the same mass distribution. In general, we found that this effect would not significantly perturb the HJ system model $\alpha$ values away from the distributions of the parent population. We tested the magnitude of the expected model modification ($\Delta\alpha$) by randomly and independently drawing the masses of two-planet systems from the derived WCJ mass distribution and discarding systems that did not produce a $3\times$ more massive outer planet. We found that the described selection process would modify the WCJ model $\alpha$ values accordingly: $\Delta\alpha\sim-0.8$ for the innermost planets and $\Delta\alpha\sim0.8$ for the outer companions. Comparatively, we found $\Delta\alpha=-0.4\pm0.3$ between HJs and WCJs, and $\Delta\alpha=0.3\pm0.5$ between HJ and WCJ companions in our observed sample. The differences in $\alpha$ trends derived for our planet sample are consistent with the expectation of our simulation to within $1.3\sigma$ for HJs and WCJ and $1\sigma$ for their respective outer companions. It is also important to note that this simple test assumes independence between planets, which is likely an oversimplification. The existence of intrasystem mass correlations would make model deviations ($\Delta\alpha$) even smaller than those reported by our simulations, given the increased likelihood of drawing a more massive outer companion. Overall, if the marginal model dissimilarities identified between HJs and WCJ systems are meaningful, these differences would likely be the result of the $3\times$ more massive outer companion requirement for HJ systems.  

\begin{figure*}
\centering \includegraphics[width=\textwidth{}]{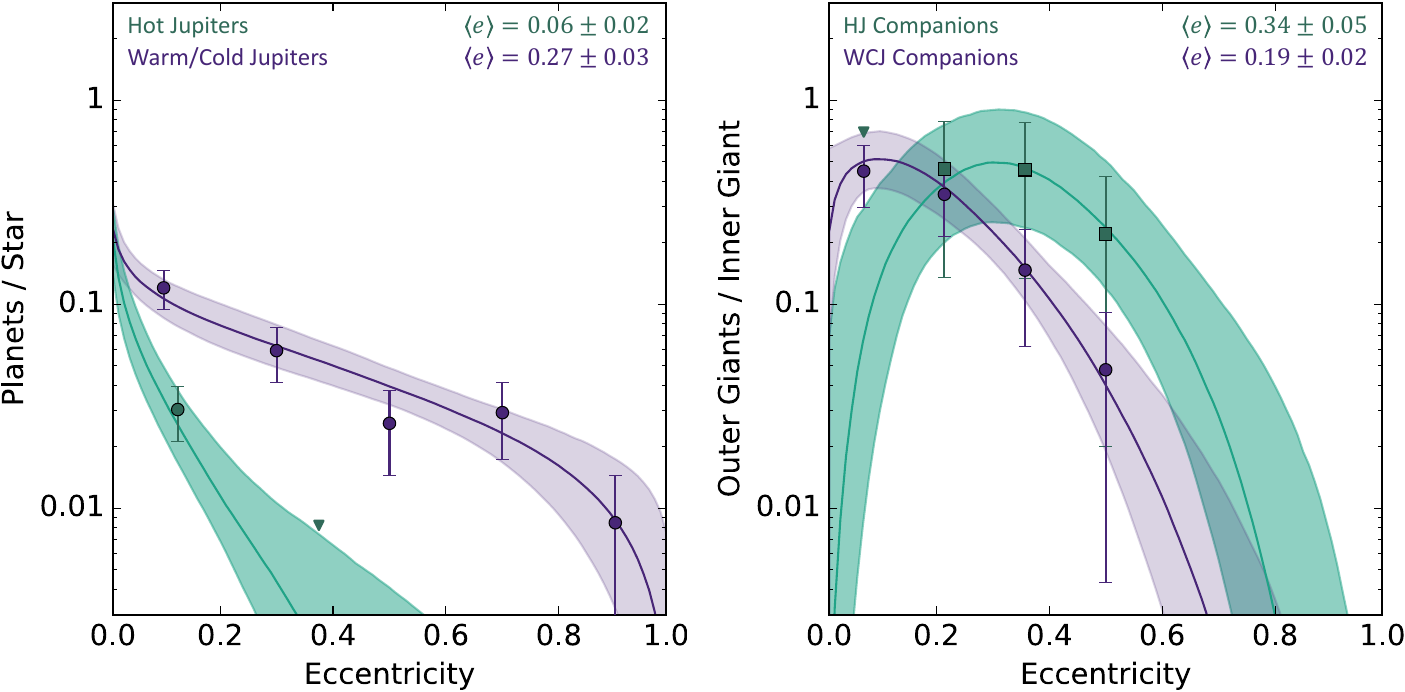}
\caption{The eccentricity distribution and the Beta distribution models for each corresponding planet population. For the hot Jupiters $\gamma=0.4\pm0.1$ and $\lambda=6.9\pm3.7$; for the warm/cold Jupiters $\gamma=0.7\pm0.2$ and $\lambda=1.8\pm0.6$; for the HJ companions $\gamma=3.7\pm1.8$ and $\lambda=7.6\pm3.4$; for the WCJ companions $\gamma=1.6\pm0.6$ and $\lambda=7.3\pm2.9$.
\label{fig:eccentricity}}
\end{figure*}

\section{Eccentricity Distribution}

The nearly circular orbits of HJs suggest a circularizing mechanism, likely involving tidal interactions. This process erases the orbital eccentricity of the inner planet, which could provide clues to the origin of HJ. However, if an outer companion exists and is identified, the eccentricity of this counterpart may retain information about the system's dynamical history.

Fig. \ref{fig:eccentricity} shows the eccentricity distributions of the four planet classes studied here\footnote{To simplify the model comparison, we rely on the expected eccentricity value for each model $\langle e\rangle=\gamma / (\gamma + \lambda)$, acknowledging that this metric fails to fully capture the width and shape of the distribution.}. As expected due to tidal circularization, the hot Jupiter population is strongly skewed towards zero $\langle e\rangle =0.06\pm0.02$. The WCJs display a wide distribution of possible eccentricities ($\langle e\rangle =0.27\pm0.03$) in agreement with \citet{kip13}, who modeled the eccentricity population of long-period giants, finding $\langle e\rangle =0.27\pm0.03$. If this distribution resembles the underlying birth population, then it is clear that most giants are born with a relatively low eccentricity, but half inherit eccentricities greater than 0.27, setting the stage for complex dynamical interactions with other existing planets.

Observing the outer giant populations can provide insight into the migration history of giant planets. Significant differences between the eccentricity distributions of inner and outer giant populations would suggest that some process modified the orbits of one or both populations. In Fig. \ref{fig:eccentricity}, we show that the expected eccentricity in the WCJ ($\langle e\rangle =0.27\pm0.03$) and WCJ companion ($\langle e\rangle =0.19\pm0.02$) distributions are different at the $2.2\sigma$ level, suggesting that some orbital interactions may have taken place. Furthermore, all the observed WCJ systems \textit{currently} have a minimum orbital separation of less than 20 mutual Hill radii (MHR; \citealt{gla93}) with a median separation of 5.9 MHR. \citet{pu15} showed that similar-mass multi-planet systems with orbital separations less than 12 MHR, and $e>0.02$, will likely become dynamically unstable within a billion years. While we do not attempt to make a claim about the stability of WCJ systems, the low MHR values imply their capacity for orbital perturbation. Thus, under the right conditions, these WCJ populations may produce a system capable of the dynamic evolution responsible for hot Jupiters.

We found that the average eccentricity of the HJ companions ($\langle e\rangle =0.34\pm0.05$) is higher than the WCJ companion population by $3\sigma$, suggesting past dynamical interactions for the HJ companions. HD 187123c, with an eccentricity of 0.23, is the lowest eccentricity HJ companion in our sample. This eccentricity enhancement for HJ companions is an expectation of high-eccentricity migration. For example, the model by \citet{petr15} predicts that after an exchange of angular momentum between two giant planets participating in this process, the outer companions will have enhanced ``mid-range'' eccentricities ($e = 0.2$--0.5).

\section{Discussion}

The WCJ and HJ populations share several demographic features, including their mass distributions, period distribution of outer giant companions, and the overall abundance of outer companions. These similarities suggest that the two populations descend from a common parent population of planets.

We have shown that having an outer companion by itself does not guarantee hot Jupiter formation. Systems of two or more WCJs are common. Our analysis, which took into account survey completeness, showed that each giant harboring star also hosts $1.0\pm0.3$ additional giants (for $M$ = 0.3--30$M_\textrm{Jup}$ and $P < 40,000$ days). However, hot Jupiters only represent $16\%$ of planets in the above mass and period range. Two cold giants can exist quiescently if the eccentricity is minimal. Such is the case for our own solar system where the low-eccentricity of both Jupiter and Saturn results in a stable secular exchange of angular momentum that will not lead to migration within the main-sequence lifetime of the Sun \citep{mur99,zin20c}.

Within our RV sample, we found that the mass distribution of hot Jupiters and warm/cold Jupiters is statistically indistinguishable. However, in systems of multiple giant planets that include an HJ, the outer giant planet has a mass that is typically $3\times$ higher. Further, we found that the outer giants in such systems are systematically more eccentric. 

The proposed criteria for hot Jupiter formation---the existence of an outer companion that is $3\times$ more massive with $e>0.2$---were determined from the observed properties of systems with HJs and WCJ companions. By applying these criteria to WCJ systems without HJs, we can estimate (without circular reasoning) what fraction of WCJ harboring systems should eventually convert one of the WCJs into an HJ. It appears that most, if not all, systems should have an outer-giant companion and that $\sim20\%$ have a mass configuration with a $3\times$ ratio. We also measured that $40\%$ of WCJs have $e > 0.2$. Thus, our simple model predicts that $\sim$10\% of systems of multiple giant planets should produce a HJ, consistent with the observed $16\pm4\%$ of giant-harboring systems that contain a HJ. This empirical population fraction was derived from the occurrence rates noted in Fig. \ref{fig:period} and was computed using:
\small \begin{equation} \label{eq:eta}
\eta_{\textrm{(HJ|giant planet)}}=\frac{f_{\textrm{HJ}}}{f_{\textrm{HJ}} + f_{\textrm{WCJ}}}=\frac{0.03}{0.03 + 0.16}=0.16\pm0.04, 
\end{equation}
where the $f$ values are population occurrence rates and $\eta_{\textrm{(HJ)}}$ is the expected fraction of giant-harboring planetary systems that contain a HJ. 

\subsection{Formation Mechanisms}

%%%%%%Kozai
Lidov-Kozai oscillations occur when two highly inclined bodies undergo a cyclical exchange of orbital eccentricity and mutual inclination. If the initial mutual inclination is sufficiently high ($i>50\degr$), the eccentricity of the inner planet can be driven to near unity. In cases where tidal disruption is avoided, the inner planet is circularized through dissipative processes during close approaches, resulting in a hot Jupiter \citep{naoz11}. These hierarchical triple systems can be either star-planet-star or star-planet-planet given the appropriate mass and semi-major axis scaling \citep{naoz16}. Because the CLS sample excludes systems with close-in stellar companions, we focused on the star-planet-planet mechanism. Lidov-Kozai oscillations produce a nearly uniform distribution of sky-projected stellar obliquities and require a more massive outer giant \citep{petr16}. Our measurements of giant-planet multiplicity and higher-mass exterior planets are consistent with this model. 

However, studies of HJ obliquities are not consistent with the Lidov-Kozai mechanism being the dominant HJ production channel.  \cite{rice22b} found that nearly 90\% of HJs are aligned with their host star's spin-axis, which is inconsistent with the expected uniform obliquity distribution for this mechanism. It may be that such the orbits of such planets have been realigned with the stellar spin axis through stellar interactions \citep{win10}. Observations of increased obliquities for hot stars ($\ge6100$ K), which should experience a prolonged re-alignment timescale, motivates such considerations. In contrast, evidence of reduced HJ occurrence for these massive stars, identified with TESS photometry \citep{bel22}, deviates from the lack of $T_\textrm{eff}$ dependence for HJ occurrence observed around FGK stars \citet{zin23}. This may indicate that higher-mass stars have a unique underlying process that is responsible for HJ production. Alternatively, the terminal orbital periods of HJs may be a function of stellar mass, truncating HJs to orbits beyond the arbitrary 1--10-day period limit. Despite these observed complexities, the initial high-inclination planetary system configuration should be rare. \citet{dong23} identified a broad zero-centered unimodal distribution of obliquities, via a hierarchical Bayesian analysis. The implications of this result remain unclear, as $28\pm9\%$ of the obliquities appear to be isotropically distributed. It may be that an existing multi-modal distribution is obscured by our limited sample and/or the underlying biases of the sample \citep{sie23}. Regardless, the clear prominence of aligned orientations---$72\pm9\%$ with stellar obliquities less than $40\degr$---provides evidence that high-initial mutual inclination mechanisms, like Lidov-Kozai oscillations, are a non-dominant formation pathway.

%%%%%In situ

In the in situ theory, HJs form close to their stars and do not migrate appreciably. The mechanism involves disks with well-tuned properties that coalesce material within an orbital radius of 1 AU \citep{bat16}. In situ formation is predicted to occur in disks with large masses. Such disks are likely to also produce multiple cold Jupiters through the core accretion mechanism in the outer disk \citep{bat23}. Our measurements are consistent with some but not all of these predicted properties of giant planet systems. We do observe that HJs usually have at least one WCJ companion. However, we do not detect a higher number of WCJs in HJ systems compared to systems without such a planet. Specifically, for HJ companions we found $1.3 \pm^{1.0}_{0.6}$ companions per HJ, suggesting that $\sim30\%$ of HJ systems have more than one outer giant. While this estimate is a lower limit, due to the inherent degeneracy between the occurrence summary statistic and system multiplicity, the $30\%$ estimate is roughly consistent with the observed one-quarter of HJ systems having a second outer companion.  

In the in situ model, HJ harboring systems arise from massive disks capable of producing an HJ and should yield numerous additional giants which are likely to experience dynamical interactions that will drive up eccentricities.
Our observation that HJ companions have higher average eccentricities than WCJ companions is consistent with this prediction. Our observed mass distributions also offer another point of comparison with the in situ model.  This theory predicts that the massive disks that produce hot Jupiters are also more likely to produce more massive giant planets in exterior orbits. This is a byproduct of the local gas-mass budget, with giants forming in the outer disk having a larger gas reservoir. However, this effect would also give rise to a similar, albeit weaker, mass correlation within the WCJ population. In the outer disk ($P>400$ days), planet mass accumulation from the natal disk is linearly dependent on the planet's Hill radius---which scales directly with the planet's orbital radius ($r$)---under the assumption of a Mestel disk profile ($S\propto1/r$). Thus, this theory predicts that more massive giant planets should be on the outside, which is inconsistent with our finding that the WCJs and their companions have very similar mass distributions (median mass ratio = 0.99). Furthermore, the underlying HJ mass distribution (Fig. \ref{fig:mass}) should appear significantly more bottom-heavy when compared to the WCJ population if the mass ordering were regulated by the local disk properties. Despite these shortcomings, \citet{mor23} showed that the in situ framework naturally reproduces the planet and heavy-element mass correlations, where more massive warm Jupiters appear with a super-stellar heavy element enrichment \citep{thor16}. The conditions in the inner disk allow for inward radial drift of dust \citep{mor23}, leading to a natural pile-up of heavy elements at short orbital periods. Thus, giants formed within 1 AU should reflect this metallicity enrichment.  While in situ formation can replicate many of the observed features, it fails to account for the lack of high multiplicity giants in the outer disk and does not currently predict the lack of a mass dependence within WCJ multiplanet systems while retaining a strong dependence in HJ systems.

% Identified by \citet{thor16}, more massive warm Jupiters appear to have a heavy element enrichment. 

% identified by \citet{thor16} for warm Jupiters. Thi Therefore, it seems possible that \textit{in-situ} formation may provide some contribution to the underlying warm Jupiter population, 

% Therefore, it seems the framework for \textit{in-situ} formation is in tension with the existing observations.

% It may be true that efficient planet-forming disks, which are necessary for \textit{in-situ} HJ formation, also produce an abundance of distant giant companions, and

% \citep{mor23} - hj fomration insitue will produce dust rich atmosphers matching observations of \citet{thor16}.

% \citep{bat23} - planets form in ring around 1 au. You need large mass buget of of rocky planetesimals, high mass buguet icy planetesmal, more cores, more runaway accretion. explains multiplcity.

%%%%%%Secular Chaos

In the secular chaos framework, multiple outer giants act to transfer angular momentum outward via overlapping resonances, exciting the eccentricity/inclination of the innermost giant \citep{wu11}. The inner planet then undergoes tidal dissipation via interactions with the host star, resulting in a hot Jupiter in a nearly circular orbit. The observables from our study mostly do not support secular chaos as the dominant mechanism for HJ production. Secular chaos requires multiple outer giant planets, and we estimate that $\sim$30\% of the systems in our sample have this characteristic\footnote{There exists a degeneracy between multiplicity and the occurrence rate $f$. In other words, it is possible to have a small fraction of systems with numerous giants and a larger number of lonely HJs, yielding equivalent results. However, a majority of the multi-planet systems in our sample appear to harbor only two planets, making this an unlikely outcome.}. Our observation that outer planets in HJ systems are more massive than their inner companions may arise naturally under the secular chaos framework, as such architectures have an increased propensity for diffusive transfers of angular momentum that help the inner planet achieve an eccentricity near unity. However, the $3\times$ mass enhancement observed for outer companions exceeds the general preference for a more massive outer campaign predicted by secular chaos. The higher multiplicity requirement of this mechanism enables lower-mass outer companions to cumulatively increase the eccentricity of the innermost giant. Furthermore, \citet{tey19} showed that the observed obliquity distribution of HJs is in tension with secular chaos, which tends to produce orbits that are moderately misaligned from the stellar rotational axes. It may be the case that many of these orbits have undergone tidal realignment \citep{rice22}, but these realignment models still over-predict the number of moderate misalignments. 

%%%%%%%High Eccentric Co-planar
Another mechanism is coplanar high-eccentricity migration \citep{petr15}, which requires an architecture of at least two giants in a nearly coplanar orbit, a planet mass ratio $M_\mathrm{outer}/M_\mathrm{inner} \gtrsim 3$, and the outermost giant having a larger eccentricity. This mechanism is the most quiescent type of high-eccentricity migration, relying on secular gravitational interactions between planets with low mutual inclinations. Initially, the system must harbor heightened eccentricities, in the form of either two planets with $e>0.5$ or a circular inner planet and an outer companion with $e>0.67$. However, if the mass of the eccentric outer planet is less than $\sim$3$\times$ the mass of the inner giant, the exchange will not significantly decrease the angular momentum of the inner planet. Upon the completion of the HJ's orbital evolution (typically 4--100 Myr.), the outer companion retains a mid-range eccentricity, which has been reduced somewhat by the preceding orbital energy exchange. Thus, the resulting system will harbor an eccentric outer giant ($e\approx0.2$--0.5) and will be $\gtrsim3\times$ more massive than the inner giant---an observable correlation that is unique to this formation pathway. After accounting for detection completeness, this is precisely what we observe. 
While six out of the eleven HJs in our sample do not have a detected outer giant, our modeled companionship rate is $1.3\pm^{1.0}_{0.6}$ planets/HJ, showing that this absence is merely the outcome of reduced detection efficiency at longer orbital periods.

Coplanar high-eccentricity migration requires that systems begin with low mutual inclinations ($i<30\degr$), making it difficult to form the complex orbital orientation seen in approximately 10\% of HJs \citep{win10}. It may be the case that torquing of the natal disk has misaligned these systems before the planets formed \citep{bat13Or}, creating coplanar systems that are misaligned from their host star's spin axis. Alternatively, a different mechanism may be at play for these systems, which tend to include more massive stars.

\section{Conclusions and Summary}
We examined the population features that make hot Jupiter (HJ) systems distinct from their warm and cold Jupiter (WCJ) counterparts, leveraging the long baseline of the California Legacy Survey RV planet catalog \citep{ros21} to identify underlying demographic similarities and differences. We found that the following characteristics are important:

\begin{itemize}
 \item Planetary systems that include a giant planet tend to have more than one such planet. For systems with an HJ, we measure $1.3\pm^{1.0}_{0.6}$ additional planets per system in a model integrated out to 40,000-day orbital periods. For WCJ systems, we find $1.0\pm0.3$ additional outer planets per system. Furthermore, we find that the distribution of orbital periods of the outer companions is similar for HJ and WCJ systems. The similarity between outer companion properties suggests that HJ systems may have started off as WCJ systems.

 \item The mass distributions of HJs and WCJs are indistinguishable. However, for systems with an HJ, the outer giant planet is typically more massive with $M_\mathrm{outer}/M_\mathrm{inner} \gtrsim 3$. For systems without an HJ, this ratio is nearly unity. This suggests that having an especially massive outer giant planet companion makes a system of multiple giant planets more likely to develop an HJ. 

 \item The eccentricity distributions of HJs, WCJs, and their companions show population-level differences. HJs are mostly circular ($\langle e\rangle =0.06\pm0.02$), as is well known. The inner giant planets in WCJ systems have higher eccentricities ($\langle e\rangle =0.27\pm0.03$) than their outer companions ($\langle e\rangle =0.19\pm0.02$). The highest eccentricities belong to the population of HJ outer companions ($\langle e\rangle =0.34\pm0.05$). We identify the presence of a more eccentric outer companion as a second factor that makes a system of giant planets more likely to develop an HJ.
\end{itemize}
Our observations suggest that the combination of these three factors creates an environment conducive to HJ formation. Under the assumption that the three factors (giant planet multiplicity, mass ratio, and outer planet eccentricity) are drawn independently, we estimate that $\sim$$10\%$ of stars hosting a giant planet should produce an HJ, which is consistent with our measurement of $16\pm4\%$. 

The patterns identified here favor coplanar high-eccentricity migration as the dominant formation mechanism for HJs. This mechanism requires a $\gtrsim 3\times$ more massive outer planet to increase the eccentricity of the inner giant planet through secular interactions, which reduce the inner planet's orbital pericenter. The orbit of the inner planet is eventually tidally circularized and reduced in radius. This mechanism is effective for systems that start with high-eccentricity outer giants, and it leaves behind a moderately eccentric relic, as seen in our HJ companion population.
The coplanar aspect of this mechanism naturally produces aligned HJ systems and would be further validated if HJ obliquity measurements continue to cluster near zero-degree orientations, as is currently the case \citep{dong23}.

\acknowledgments
JZ and AH thank Lee Rosenthal, BJ Fulton, Konstantin Batygin, Cristobal Petrovich, Erik Petigura, Heather Knutson, and Steven Giacalone for their thoughtful insights and feedback on this work. We also thank the anonymous referee for their insightful comments. 

JZ acknowledges support provided by NASA through the Hubble Fellowship grant HST-HF2-51497.001 awarded by the Space Telescope Science Institute, which is operated by the Association of Universities for Research in Astronomy, In., for NASA, under the contract NAS 5-26555. 

\bibliography{paper.bbl}{}
\bibliographystyle{aasjournal}

\end{document}